\documentclass[pra,twocolumn,showpacs,nobibnotes,floatfix,superscriptaddress]{revtex4}

\usepackage{latexsym,amsmath,amssymb,amsfonts,mathbbol,graphicx,color}
\usepackage{dcolumn}
\usepackage{bm}

\begin{document}

\title{Tri-partite Entanglement Witnesses and Sudden Death}

\author{Yaakov S. Weinstein}
\affiliation{Quantum Information Science Group, {\sc Mitre},
260 Industrial Way West, Eatontown, NJ 07224, USA}


\begin{abstract}
I explore entanglement dynamics in a three qubit system comparing the ability of 
entanglement witnesses to detect tri-partite entanglement to the phenomenon of 
entanglement sudden death (ESD). Using a system subject to dephasing I invoke 
entanglement witnesses to detect tri-partite GHZ and W-type
entanglement and compare the evolution of their detection capabilites with the 
evolution of the negativity, bi-partite concurrence, and  tri-partite negativity. 
Interestingly, I find a state in which there is no concurrence or tri-partite 
negativity but there is entanglement. Finally, I utilize a 
three qubit quantum error correction (QEC) code to address how ESD affects 
the abilities of quantum error correction.
\end{abstract}


\maketitle

\section{Introduction}

Entanglement is a uniquely quantum mechanical phenomenon in which quantum 
systems exhibit correlations not possible for classical systems. As such, 
entanglement is a vital resource for many aspect of quantum information processing 
including quantum computation, quantum metrology, and quantum communication \cite{book}.
But despite its fundamental and practical importance and much work in the subject,
there are many aspects of entanglement, especially multi-partite entanglement,
that are in need of further study \cite{HHH}.

A major challenge facing experimental implementations of quantum computation, sensing,
and communication is decoherence, unwanted interactions between the system
and environment. Decoherence may be especially detrimental to highly non-classical,
and hence most potentially useful, entangled states \cite{Dur}. A manifestation 
of this is entanglement suddent death (ESD) in which entanglement is completely lost
in a finite time \cite{DH,YE1} despite the fact that the coherence loss of the system 
is asymptotic. This aspect of entanglement has been well explored in the case of 
bi-partite systems and there are a number of studies looking at ESD in multi-partite 
systems \cite{SB,ACCAD,LRLSR,YYE}. In addition, there have been several initial 
experimental studies of this phenomenon \cite{expt}. However, even when analyzing a 
multi-partite system, previous works demonstrate
ESD only for bi-partite entanglement, either via concurrence or negativity 
rather than using measures for purely multi-partite entanglement. It is important 
to note that the characterization and quantification of true multi-partite entanglement 
is still very much an unsettled area for pure states and even more so for mixed states. 

In this paper I explore the loss of detectable entanglement in a three qubit system 
by invoking entanglement witnesses, observables that can detect the presence of 
entanglement. Entanglement may be present in a system but still not be practically useful 
\cite{ACCAD}. For entanglement to be useful its presence should be efficiently detectable 
experimentally. Multi-partite entanglement can be detected inefficiently via quantum state 
tomography or a violation of Bell inequalities. It can be detected efficiently by utilizing 
properly constructed entanglement witnesses \cite{TG}. 
I compare the entanglement detection abilites of tri-partite entanglement witnesses 
to ESD of tri- and bi-partite entanglement in a given system. 
In this exploration I find a state which has no concurrence and no tri-partite entanglement as 
measured by the tri-partite negativity but is entangled as measured by the 
negativity. I then apply these results to 
a three qubit quantum error correction (QEC) code and explore how ESD affects the working 
of this code. 
 
Three qubit pure states can assume a `standard' form with respect to local 
unitary operations \cite{AACJLT},
\begin{equation}
|\psi\rangle = \lambda_0|000\rangle+\lambda_1e^{i\theta}|100\rangle+
\lambda_2|101\rangle+\lambda_3|110\rangle+\lambda_4|111\rangle,
\end{equation}
where $\lambda_i > 0$, $\sum_i\lambda_i^2 = 1$ and $0 \leq \theta \leq \pi$.
These states can be separated into four broad categories: separable (in all three qubits), 
biseparable, and there exist two types of locally inequivalent tri-partite 
entanglement (GHZ and W-type) \cite{DVC}. Similar classification schemes exist for mixed 
states. Reference \cite{ABLS} defines four classes of three qubit mixed states each of which 
includes the preceeding classes as special cases.
They are separable (S) states, bi-separable (B) states, W states, and GHZ states, which 
encompasses the complete set of three qubit states. Note that additional subtlety exists in 
characterizing the entanglement within each of these classes \cite{SGA}. 

To determine in which class a given state belongs one can use entanglement 
witnesses, observables which give a positive or zero expectation value for 
all states of a given class and negative expectation values for at least one 
state in a higher (i.e. more inclusive) class. Specifically, I will make use of 
entanglement wintesses \cite{ABLS} to identify whether a state is in the 
GHZ$\backslash$W class (i.~e.~ a state in the GHZ class but not in the W class), 
in which case it certainly has GHZ type tri-partite entanglement, 
the W$\backslash$B class, in which case the state certainly has true tri-partite entanglement 
either of the GHZ-type or W-type, or the B class in which case it is not certain that the state 
has any tri-partite entanglement. While the witnesses we explore may not be of the sort that 
can be implemented efficiently for experimentally determining the 
presence of tri-partite entanglement, they are among the most sensitive, or finest, 
known witnesses. Thus, if these witnesses do not detect the presence of entanglement 
neither will any of the effeciently implementable witnesses.

I will compare the detection ability of the witnesses to the evolution of the  
concurrence \cite{conc}, $C_{jk}$, for measuring the bi-partite entanglement 
between qubits $j$ and $k$ after partial trace over one qubit. The concurrence 
of a two qubit state with density matrix $\rho_{jk}$ is defined as the maximum 
of zero and $\Lambda$, where 
$\Lambda = \sqrt{\lambda_1}-\sqrt{\lambda_2}-\sqrt{\lambda_3}-\sqrt{\lambda_4}$
and the $\lambda_i$ are the eigenvalues of 
$\rho_{jk}(\sigma_y^j\otimes\sigma_y^k)\rho_{jk}^*(\sigma_y^j\otimes\sigma_y^k)$
in decreasing order, where $\sigma_y^i$ is the $y$ Pauli matrix of qubit $i$. 
Other entanglement measures that I will look at are the negativity, $N$, for which I will use 
the sum of the absolute values of the negative eigenvalues of the partial transpose 
of the density matrix \cite{neg} with respect to one qubit, and the tri-partite negativity, 
$N_3$, a tri-partite entanglement 
measure for mixed states which is simply the third root of the product of the 
negativities with respect to each of three qubits \cite{SGA}. If the negativity
is the same when taking the partial transpose with respect to any of the three qubits,
$N = N_3$. 

We look at a three qubit system, with no interaction between the 
qubits, placed in a dephasing environment fully described by the Kraus operators
\begin{equation}
K_1 = \left(
\begin{array}{cc}
1 & 0 \\
0 & \sqrt{1-p} \\
\end{array}
\right); \;\;\;\;
K_2 = \left(
\begin{array}{cc}
0 & 0 \\
0 & \sqrt{p} \\
\end{array}
\right)
\end{equation} 
where the dephasing parameter $p$ can also be written in a time-dependent fashion 
$p = 1-\exp(-\kappa t)$. When all three qubits undergo dephasing we have eight 
Kraus operators each of the form 
$A_l = (K_i\otimes K_j\otimes K_k)$ where $l = 1,2,...,8$ and $i,j,k = 1,2$.

The next three sections are dedicated to exploring the entanglement dynamics 
and detectability of three types of three qubit states: GHZ-type states in 
Section \ref{Sghz}, W-type states in Section \ref{Sw}, and a state with 
GHZ-type tri-partite entanglement and bi-partite entanglement in Section
\ref{Sgb}. In Section \ref{Sqec} we apply our results to the three qubit 
phase flip quantum error correction code and study the affect of ESD on the 
code. Section \ref{Sconc} contains some further observations and conclusions.

\section{GHZ-type States}
\label{Sghz}

In this section I explore the affects of dephasing on the following 
three qubit GHZ-type states, 
\begin{equation}
\rho = \frac{1-q}{8}\openone+q\rho_{G}
\end{equation}
where $\rho_G = (|000\rangle+|111\rangle)(\langle000|+\langle111|)$.
These states have no bi-partite entanglement and non-zero negativity and 
tri-partite negativity equal to $-\min\left[\frac{1}{8}(1-5q),0\right]$.
The state $\rho$ is certainly in the GHZ$\backslash$W class for 
$q > \frac{5}{7}$ since it can be detected via the witness \cite{ABLS}
$\mathcal{W}_G = \frac{3}{4}\openone-\rho_G$. $\rho$ is in at least the 
class W$\backslash$B for $q > \frac{3}{7}$ since it can be detected 
via the witness \cite{ABLS} $\mathcal{W}_{W2} = \frac{1}{2}\openone-\rho_G$. 

The effect of dephasing all of the qubits of $\rho$ 
is simply to reduce the magnitude of the off diagonal elements 
$\frac{q}{2}\rightarrow \frac{q}{2}e^{-3\kappa t/2}$. For $q = 1$ 
the entanglement witness $\mathcal{W}_{W2}$ gives 
$-\frac{1}{2}e^{-3\kappa t/2}$ which decays exponentially 
with time. In other words, the state can always be detected by a 
W-state witness and does not exhibit tri-partite ESD \cite{ACCAD}. 
However, the dephased state cannot always be detected as belonging to
the GHZ$\backslash$W class. This can be seen from the expectation value
$\rm{Tr}(\mathcal{W}_G\rho)$ which is negative (thus detected) only as long as 
$\kappa t \leq \frac{2}{3}\ln(2)$. Thus, we have a `sudden death' type phenomenon
with respect to the GHZ entanglement witness. Experimentally we cannot be sure the 
state exhibits GHZ-type entanglement.

\begin{figure}[t]
\includegraphics[width=4.25cm]{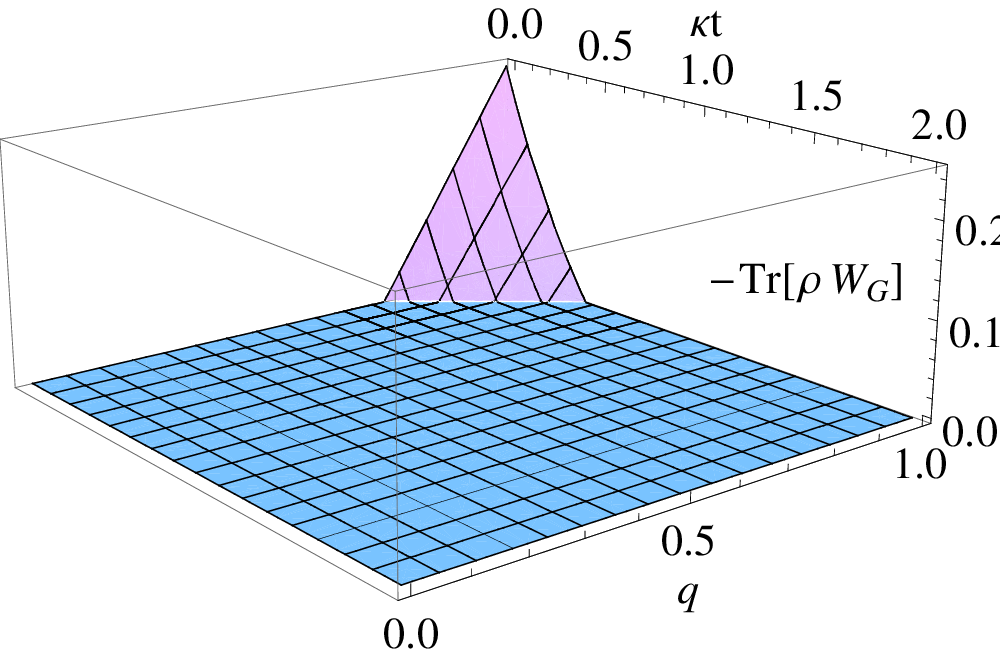}
\includegraphics[width=4.25cm]{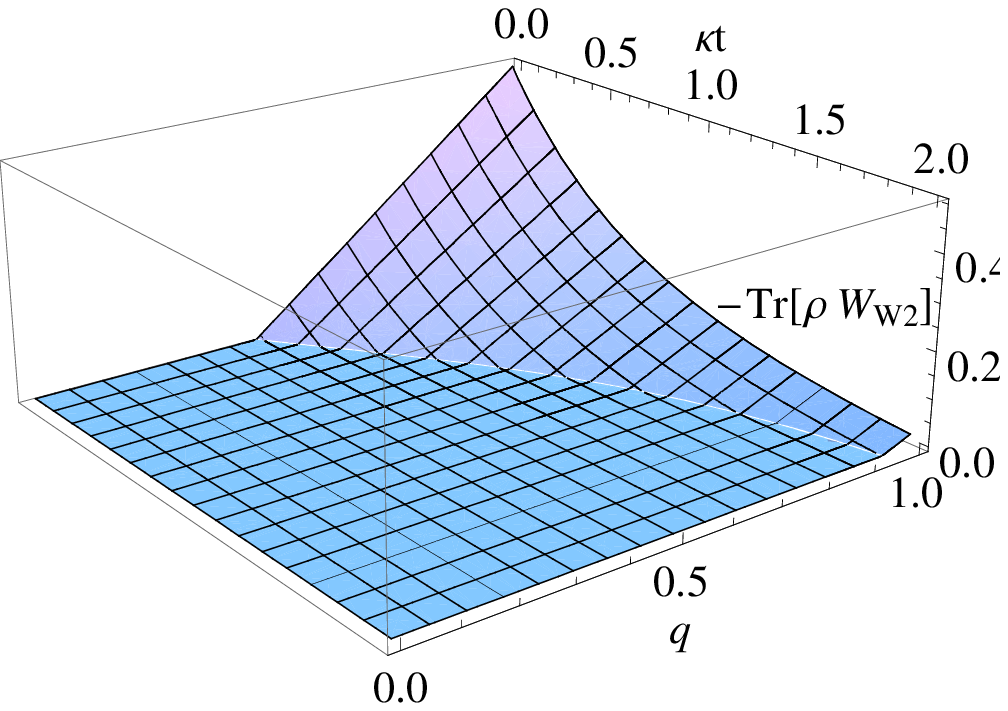}
\includegraphics[width=4.25cm]{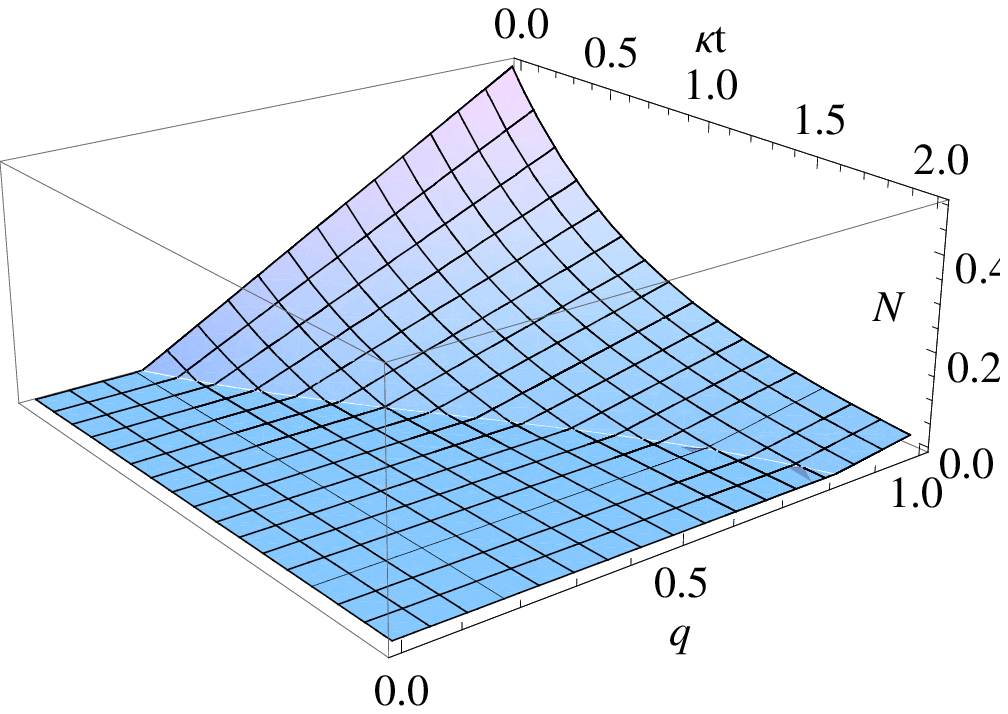}
\caption{(Color online) Demonstration of tri-partite entanglement sudden death in GHZ states. 
Top-left: Expectation value of the entanglement witness $-\rm{Tr}(\mathcal{W}_{G}\rho)$ 
as a function of the initial state
paramaterized by $q$, and the dephasing constant $\kappa t$. At sufficiently low
values of $q$ or high values of $\kappa t$ the state can no longer be detected 
by a GHZ-type entanglement witness. 
Top-right: $-\rm{Tr}(\mathcal{W}_{W2}\rho)$ as a function of the same parameters. 
This entanglement witness shows when the state can no longer be detected as one 
with tri-partite entanglement.
Bottom: The negativity with respect to any qubit partition (and thus in this case 
the negativity is equivalent to the tri-partite negativity). When this is 
zero the state has no distillable entanglement. For $q = 1$ (initial pure state) 
the negativity is equal to $-\rm{Tr}(\mathcal{W}_{W2}\rho)$ meaning that 
the $\mathcal{W}_{W2}$ entanglement witness always detects the presence of the
tri-partite entanglement. When $q < 1$ the state is mixed and the witness does
not always detect the tri-partite entanglement which is known to be present via the 
tri-partite negativity measure.
}
\label{GHZstate}
\end{figure}

The inability to detect any tri-partite entanglement in the state $\rho$ 
can occur for $q < 1$. In such a case we find 
$\rm{Tr}(\mathcal{W}_{W2}\rho) = \frac{1}{8}(3-3q-4qe^{-3\kappa t/2})$ and
$\rm{Tr}(\mathcal{W}_G\rho) = \frac{1}{8}(5-3q-4qe^{-3\kappa t/2})$. 
The behavior, as a function of time, for both entanglement witnesses is 
similar, as shown in Fig.~\ref{GHZstate}, and clearly demonstrates the 
inability of the witnesses to detect any entanglement at sufficiently long
but finite time. 

Comparing this to the entanglement measures, we note that there is only 
one eigenvalue of the partially transposed density matrix that can be negative and so 
$N = N_3 = -\min\left[\frac{1}{8}(1-q-4qe^{-3\kappa t/2}),0\right]$. 
For $q = 1$, i.e. initially pure states, this is equal to 
$-\rm{Tr}(\mathcal{W}_{W2}\rho)$ and thus the witness always detects the 
tri-partite entanglement. For $q < 1$ the initial state is mixed
and $-\rm{Tr}(\mathcal{W}_{W2}\rho)$ is no longer equal to $N$. 
Instead, the entanglement witness goes to zero
at a rate approaching three times that of $N$ demonstrating that there is
remaining tri-partite entanglement in the system beyond the detection of 
the entanglement witness. Nevertheless, even $N$ decays 
to zero in finite time exhibiting the ESD phenomenon, Fig.~\ref{GHZstate}.
As expected, $C$ between any two qubits of $\rho$ after partial trace over 
the third qubit is always zero thus the system never exhibits bi-partite 
entanglement. 

\section{W-type States}
\label{Sw}

Unlike GHZ states the W state 
$|W\rangle = \frac{1}{\sqrt{3}}(|001\rangle+|010\rangle+|100\rangle)$, 
retains a high degree of (bi-partite) entanglement upon partial trace 
over one qubit. Thus, W-type states allow some comparison between 
tri-partite and bi-partite entanglement evolution. We start with the state
\begin{equation}
\rho_ = \frac{1-q}{8}\openone+q\rho_{W}.
\end{equation}
where $\rho_W = |W\rangle\langle W|$.
States of this sort in the W$\backslash$B class may be detected by the 
entanglement witness \cite{ABLS} $\mathcal{W}_{W1} = \frac{2}{3}\openone-\rho_W$. 
For $q = 1$ the expectation value of the entanglement witness is 
$-\frac{1}{3}$ and the concurrence of any two qubits is $\frac{2}{3}$.
The entanglement witness detects the state, for all 
$q > \frac{13}{21} \simeq .619$. $C$, after partial trace of a
qubit, remains positive for all $q \agt .5482$ and $N$, which 
is equal to $N_3$ since the results of the partial trace over 
any of the three qubits are equivalent, is non-zero for $q > .2096$. 
Once again the witness is successful in detecting tri-partite entanglement
only up to a certain point despite the continued presence of tri-partite 
entanglement. In addition, for $.2096 \alt q \alt .5482$ there is surviving 
tri-partite $W$-type entanglement, as measured by $N_3$, even when there is no 
longer any residual bi-partite entanglement.

\begin{figure}[t]
\includegraphics[width=4cm]{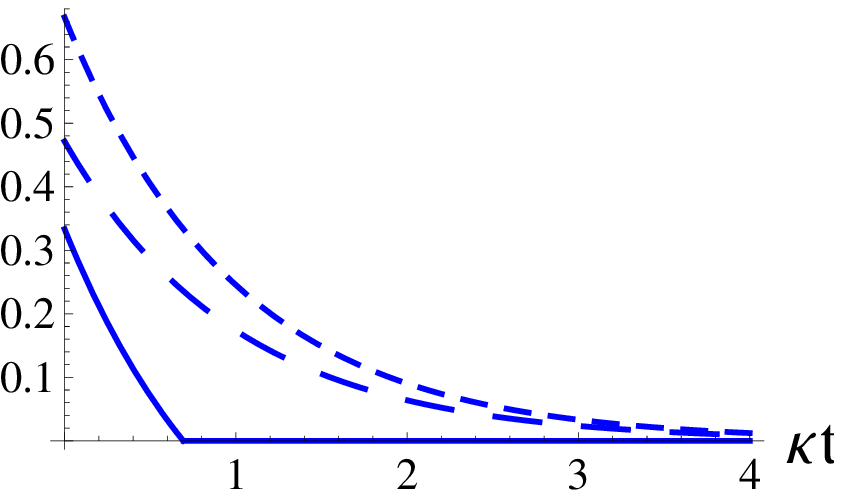}
\includegraphics[width=4cm]{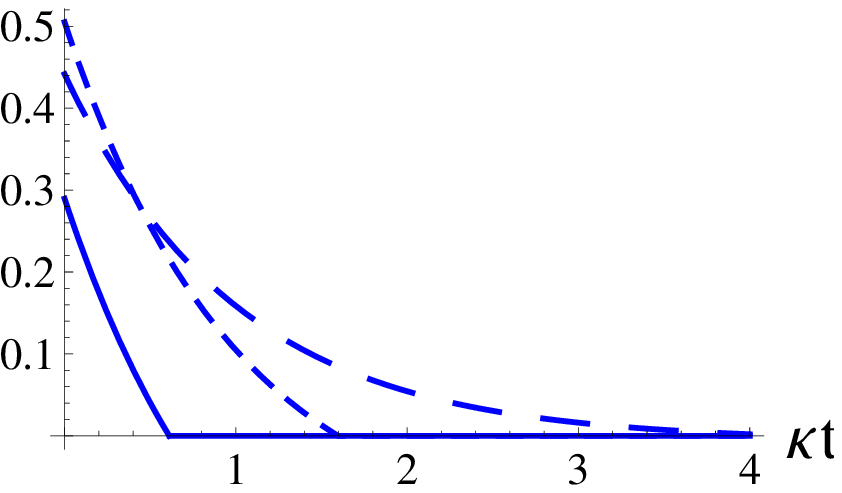}
\includegraphics[width=4cm]{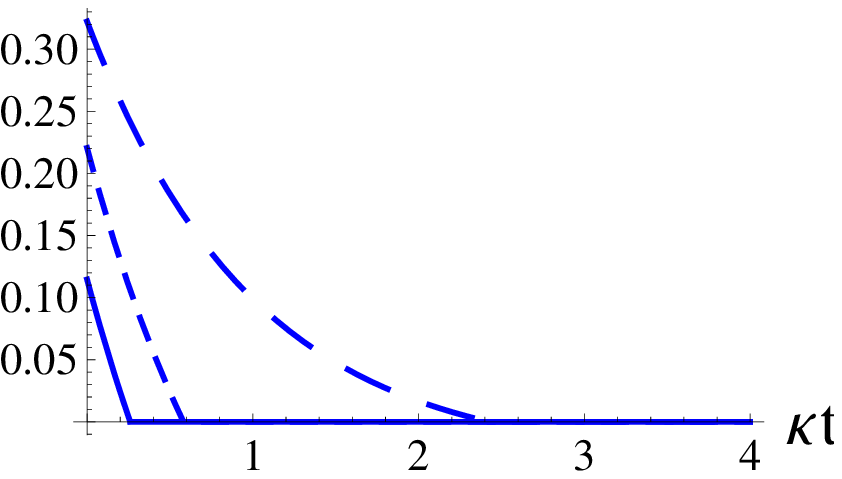}
\caption{(Color online) Demonstration of bi- and tri-partite entanglement sudden 
death in W-states. Each plot shows the expectation value of the 
entanglement witness $-\rm{Tr}(\mathcal{W}_{W}\rho_W)$ (solid line),
the bi-partite concurrence (medium dashed line), and the negativity
(large dashed line) as a function of the dephasing constant $\kappa t$. 
Top left: $q = 1$, the initial state is pure. In this case 
$-\rm{Tr}(\mathcal{W}_{W}\rho_W)$ goes to zero in finite time, 
but the other entanglement measures do not. 
Top right: $q = .95$, for any $q < 1$ the concurrence undergoes ESD
before the negativity (which is equal to the tri-partite negativity) 
goes to zero. Bottom: $q = .75$.}
\label{Wstate}
\end{figure}

The effect of dephasing on the state $\rho$ is to degrade the off-diagonal
terms by $e^{-\kappa t}$. The state is detected by $\mathcal{W}_{W1}$ only
for $\kappa t < \ln16-\ln(-5+\frac{13}{q})$ while $C$, after tracing out
any qubit, is $\max\left[\frac{1}{6}\right(4e^{-\kappa t}q-\sqrt{-3(q-1)(q+3)}\left),0\right]$. 
We compare these values to $N$ where again there is only one negative 
eigenvalue upon partial transpose of the density matrix: 
$N = -\min\left[\frac{1}{24}(3-q(3+8\sqrt{2}e^{-\kappa t})),0\right]$.
For $q = 1$ and sufficiently large $\kappa t$ we do not find ESD
for $C$ or $N$ but nevertheless find that the tri-partite entanglement 
cannot be detected by $\mathcal{W}_{W1}$. For 
$q < 1$ ESD occurs for $C$ and $N$ as shown in Fig.~\ref{Wstate}
with the tri-partite entanglement witness going to zero first followed by $C$
and finally the $N = N_3$. In other words, after a certain time there is no
longer bi-partite entanglement in the system despite the remaining W-type 
tri-partite entanglement. The above expressions also show that each entanglement measure 
decreases at the exponential rate $-\kappa t$ and there is no difference in the 
exponential rate of decay between the bi-partite and tri-partite entanglement.
The difference in entanglement measure evolution comes from different coefficients mulitplying the 
exponential and additional constants.

\section{GHZ-type State with Bi-Partite Entanglement}
\label{Sgb}

We would like to explore a system that contains both bi-partite and 
tri-partite GHZ-type entanglement to compare and contrast the two 
types of entanglement evolution. We look at the following initial state
\begin{equation}
\rho_ = \frac{1-q}{8}\openone+q\rho_{gb},
\end{equation}
where $\rho_{gb} = |\psi\rangle_{gb}\langle\psi|$ and
$|\psi\rangle_{gb} = \frac{2}{3}(|000\rangle+|111\rangle)+\frac{1}{3}|110\rangle$.
In order to detect the tri-partite entanglement of such states we must first
find the proper GHZ and W-type entanglement witnesses. The closest
state to $|\psi\rangle_{gb}$ with no W-type vectors is likely to be the GHZ state
with an $x$-rotation about the third qubit. With this in mind we construct GHZ
and W state witnesses $\mathcal{W}_{Ggb} = \frac{3}{4}\openone-\rho_G(\theta)$ and 
$\mathcal{W}_{Wgb} = \frac{1}{2}\openone-\rho_G(\theta)$, for third qubit rotation 
of state $\rho_G$ by angle $\theta$. The $\theta$ that minimizes the witnesses is 
$\theta \simeq 28.075^{\circ}$.

With these tools we can explore the entanglement of this system. With
no dephasing the state $\rho$ is detected by $\mathcal{W}_{Ggb}$ for 
$q \agt .763$ and detected by $\mathcal{W}_{Wgb}$ for $q \agt .458$.
Taking a partial trace over one of the qubits leads to non-zero 
concurrence only when tracing over the third qubit. This
concurrence, $C_{12}$, is non-zero only when $q \agt .529$. The 
lowest lying eigenvalue of the partial transpose of $\rho_{gb}$ is when 
the partial transpose is taken with respect to the first or second qubit
and gives non-zero negativity for $q \agt .201$. The tri-partite negativity
in this case is not equal to the negativity and is nonzero only 
for $q \agt .220$. 

\begin{figure}[t]
\includegraphics[width=4cm]{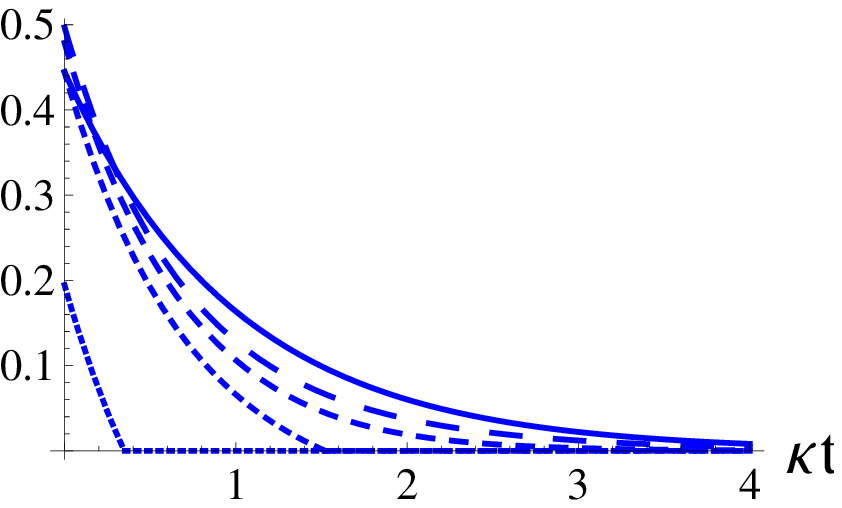}
\includegraphics[width=4cm]{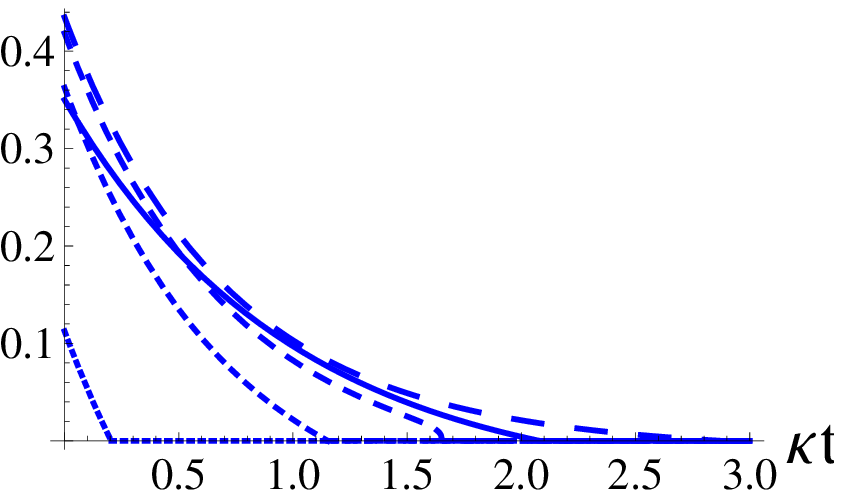}
\includegraphics[width=4cm]{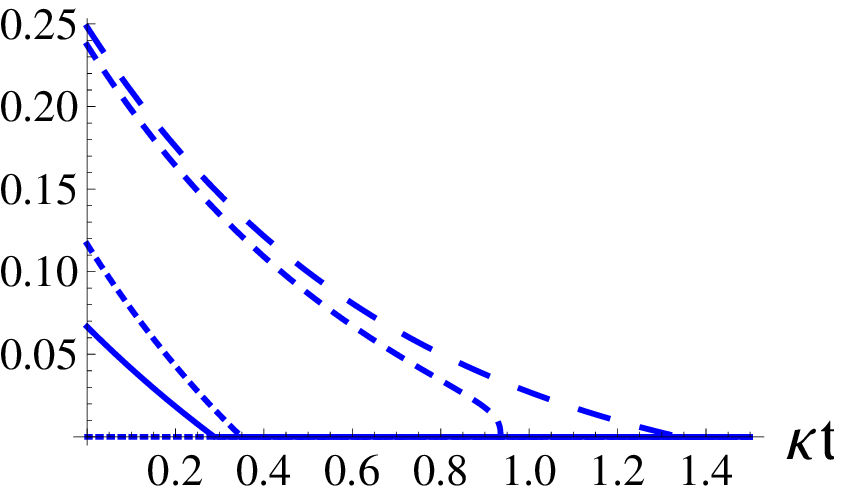}
\caption{(Color online) Demonstration of entanglement sudden 
death in $\rho_{gb}$-type states. Each plot shows $C_{12}$ (solid line), $N$
(large dashed line), $N_3$ (medium dashed line), $-\rm{Tr}(\mathcal{W}_{Ggb}\rho_{gb})$ 
(small dashed line), and $-\rm{Tr}(\mathcal{W}_{Wgb}\rho_{gb})$ (dotted line)
as a function of the dephasing constant $\kappa t$. 
Top left: $q = 1$, the initial state is pure. In this case the 
tri-partite entanglement witnesses go to zero in finite time 
but the concurrence, negativity, and tri-partite negativity do not
undergo ESD. 
Top right: $q = .9$, for any sufficiently high value of $q$ where 
$q < 1$ the concurrence undergoes ESD
before the negativity but after the tri-partite negativity.  
Bottom: $q = .6$, the concurrence undergoes ESD before the 
tri-partite negativity and even before the tri-partite entanglement
witness. Thus, there is detectable tri-partite entanglement in the 
system though there is no concurrence remaining when one of the qubits
is traced over.}
\label{GHZBstate}
\end{figure}

The effect of dephasing on the entanglement of this state is shown in Fig.~\ref{GHZBstate}. 
For $q = 1$ the tri-partite entanglement witnesses cannot detect the entanglement
below a certain threshold $\kappa t$, but $C_{12}$, $N$,and $N_3$ do not exhibit ESD. 
For $q < 1$ all of the entanglement measures exhibit ESD. 
It is interesting to compare and contrast the behavior of the various entanglement 
measures especially with respect to $C_{12}$.
When $q < 1$ the entanglement that takes the longest amount of time to 
exhibit ESD is $N$. For sufficiently high values of $q$ the tri-partite
negativity goes to zero before $C_{12}$. For the time between the subsequent
sudden death of $C_{12}$ and the ESD of $N$ the entanglement present in the 
system is not of the bi-partite type since it is not destroyed by the partial 
trace, nor is it tri-partite in the sense that it can be measured by $N_3$.
The ability of the tri-partite entanglement witnesses to detect entanglement
is weaker than any of the entanglement measures. 

As $q$ decreases $C_{12}$ decreases at a faster rate than any of the other 
entanglement measures. 
For sufficiently low values of $q$ the concurrence, $C_{12}$, exhibits ESD 
before $N_3$ such that the bi-partite entanglement is the first to experience
ESD, followed by the tri-partite entanglement measured by $N_3$. After this time 
there is entanglement that can be measured only by $N$. The behavior of $N_3$ is 
different than that of the other entanglement measures. Rather than an exponential
decay minus some constant there is initial exponential decay and then an inflection point
before $N_3$ goes to zero.
Decreasing $q$ futher we see that $C_{12}$ undergoes ESD even before the W-state 
entanglement witness can no longer detect entanglement. This is 
in sharp contrast to the W-state case in which $\rm{Tr}(\mathcal{W}_{W1}\rho)$
always went to zero before the bi-partite $C$.
This again shows the possibility of states with tri-partite W-type entanglment 
without the presence of $C$ when tracing over one of the qubits.

\begingroup
\squeezetable
\begin{table*}
\caption{Summary of tri-partite entanglement decay behavior for GHZ-type,
W-type, and states with GHZ and bipartite entanglement (GB) for expectation values
of entanglement witnesses detecting tri-partite entanglement (the expectation value
is negative when entanglement is detected), the negativity, the tri-partite negativity 
and the concurrence between the first and second qubits (all of which are the maximum 
between the values in the table or zero). }
\begin{tabular}{||c||c|c|c||}
\hline 
 & GHZ & W & GB \\\hline
\hline
$\mathcal{W}$ & $-\frac{1}{8}\left(3-3q-4qe^{-3\kappa t/2}\right)$ & $-\frac{1}{24}\left(13-q\left(5+16e^{-\kappa t}\right)\right)$ & $-\frac{1}{72}\left(27-q\left(9+14\cos\theta+8e^{-\kappa t}\sin\theta+8e^{-3\kappa t/2}(2+2\cos\theta+e^{\kappa t}\sin\theta)\right)\right)$\\\hline
$N$    	& $-\frac{1}{8}\left(1-q-4qe^{-3\kappa t/2}\right)$  & $-\frac{1}{24}(3-q(3+8\sqrt{2}e^{-\kappa t}))$ & $-\frac{1}{72}\left(-9+9q+16q\sqrt{e^{-3\kappa t}\left(4+e^{\kappa t}\right)}\right)$\\\hline
$N_3$  	& $-\frac{1}{8}\left(1-q-4qe^{-3\kappa t/2}\right)$ & $-\frac{1}{24}(3-q(3+8\sqrt{2}e^{-\kappa t}))$ & no analytic solution found \\\hline
	& & & $\frac{q-1}{2}-$\\ $C_{12}$ & 0 & $\frac{1}{6}\left(4qe^{-\kappa t}-\sqrt{-3(q-1)(q+3)}\right)$ & $\frac{1}{36}\sqrt{81+q^2\left(77+64e^{-2\kappa t}\right)+2q\left(81-8e^{-\kappa t}\sqrt{(9+7q)(9+11q)}\right)}+ $\\ & & & $\frac{1}{36}\sqrt{81+q^2\left(77+64e^{-2\kappa t}\right)+2q\left(81+8e^{-\kappa t}\sqrt{(9+7q)(9+11q)}\right)}$    \\
\hline
\end{tabular}
\label{Tab1}
\end{table*}
\endgroup

The results of the above explorations are summarized in the Table \ref{Tab1}. There are 
a few points worth noting. First, we see the shortcomings of the entanglement witnesses
in detecting mixed state entanglement. This is especially important for experimental 
realizations where entanglement witnesses may suggest there is no entanglement present 
between systems when in fact there is. Second, for the GHZ and W-type states the different
entanglement measures decrease at the same rate, {\emph{i.~e.~}}, the exponential term is
the same. However, this is not so for the state that has both GHZ and bi-partite entanglement.
Finally, we have found states where the negativity is non-zero despite the states having 
no tri-partite negativity and no concurrence between any set of two qubits. 

\section{Application to Quantum Error Correction}
\label{Sqec}

With what we have learned concerning the entanglement behavior of tri-partite systems
and the advent of ESD, we address the question of whether ESD affects the reliability of 
quantum error correction (QEC) or can ESD be used as a signature that error 
correction has only been successful up to a certain accuracy. Initial explorations of 
a possible connection were reported in Ref.~\cite{QECESD}. Due to the fragility 
of certain quantum states QEC will be a necessary component of any working 
quantum computer. The multi-qubit logical states in which quantum information is 
encoded by a QEC code are typically highly entangled at the level of the physical 
qubits and thus may be subject to ESD. I explore the reliability of a qubit of 
quantum information encoded into a QEC code capable of correcting phase flips
on one (physical) qubit. The goal is to determine whether ESD is a reliable signature to the 
success or failure of the code.

To encode the state $|\psi\rangle = \alpha|0\rangle+\beta|1\rangle$ into the 
phase flip code one simply uses the unencoded qubit as the control of 
two controlled-NOT gates each gate implemented between itself and an additional 
qubit in state $|0\rangle$. This is followed by Hadamard gates on all qubits. 
For simplicity we have chosen $\alpha, \beta \in \Re$. Unless $\alpha = 1$ or $\beta = 1$ 
there is always some tri-partite entanglement 
present in this state which can be detected by a proper entanglement witness and 
measured by $N$ (which in this case is equivalent to $N_3$). 
We use the follwing W-state witness to detect all types of tri-partite entangement
$\mathcal{W}_{WH} = \frac{1}{2}\openone-H\rho_{G}H$ where $H$ is a Hadamard 
on all three qubits. Detection of a possible error is done via syndrome measurement. 

When only one of the three physical qubits undergoes dephasing 
there are parameters of the intial state $|\psi\rangle$ and the dephasing parameter,
$p$, which lead to finite time non-detection of entanglement in the error state $\rho_{e1}$ 
by the entanglement witness $\mathcal{W}_{WH}$. However, there is no ESD
of $N$ or $N_3$ (though $N \neq N_3$). The three qubit QEC code always 
corrects the error. Thus, there is no clear correlation between the detection of 
tri-partite entanglement by the entanglement witness or any of the entanglement 
measures and the success of the QEC. 

\begin{figure*}[t]
\includegraphics[width=16cm]{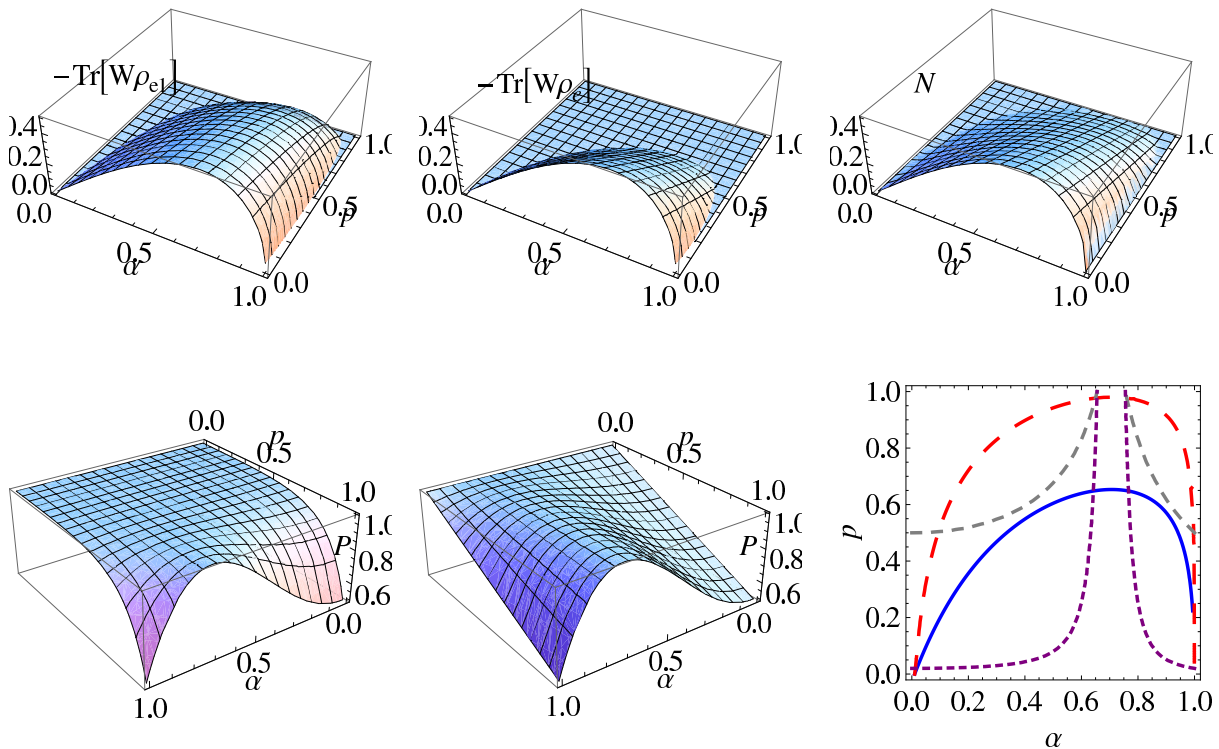}
\caption{(Color online)
Top left: Expectation value of entanglement witness 
$-\rm{Tr}(\mathcal{W}_{WH}\rho_{e1})$ for single qubit dephasing
error strength parameter $p$, before syndrome measurement and recovery, 
as a function of $p$ and the initial state of the unencoded qubit 
parameterized by $\alpha$. Lack of detection by the entanglement witness 
is evident despite the fact that error correction always works.
Top center: $-\rm{Tr}(\mathcal{W}_{WH}\rho_{e})$ with dephasing of strength
$p$ on all three qubits. Lack of detection occurs at much lower values of $p$ 
than when dephasing affects only one of the qubits.
Top right: Negativity (which is equal to $N_3$) of $\rho_e$ with dephasing of 
all qubits. ESD of negativity requires much higher values of $p$ than lack of 
detection by the tri-partite entanglement witness.      
Bottom left: Purity, $P$, of final single qubit state after error correction 
and decoding when the syndrome measurement does not detect an error. 
Bottom center: $P$ of final single qubit state after error correction 
and decoding when the syndrome measurement does detect an error.
Bottom right: Comparison of $P$ and the expectation value of the entanglement witnesses
as a function of the initial state, paramaterized by $\alpha$, and the dephasing 
strength, $p$. The contours are for $-\rm{Tr}(\mathcal{W}_{WH}\rho_e) = .001$ (solid line),
$N = .001$ (long dashed line), $P = .99$ if an error was not detected (medium dashed line), 
and $P = .99$ if an error was detected (small dashed line). The shape of the curves 
for the entanglement witness and $N$ are similar though the entanglement witness fails to 
detect the entanglement significantly before ESD of $N$. However, while areas of higher 
purity and more entanglement are at lower values of $p$ and values of $\alpha$ closer to 
$1/\sqrt{2}$, there is no apparent correlation between the presence of entanglement and 
the success of the QEC code.
}
\label{QEC}
\end{figure*}

When all three qubits are placed in a dephasing environment (where, for simplicity
we have assumed that the dephasing parameter $p$ is the same for all three qubits 
and thus ensure that $N = N_3$) the density matrix after the dephasing, $\rho_e$, exhibits 
tri-partite ESD, 
as shown in Fig.~\ref{QEC}. Moreover, the QEC code will not, in general, completely 
protect the encoded quantum information. We can measure
the effectiveness of the QEC code by looking at the purity, $P = {\rm{Tr}}(\rho_f^2)$ of the 
final single qubit state, $\rho_f$ after error correction and decoding or by looking 
at the fidelity of $\rho_f$ to the initial state. For the case at hand these two measures exhibit 
almost the exact same behavior. We can then ask whether the parameters for which 
this indicator is minimized (maximized) are those which exhibit (do not exhibit) ESD. 
We note that in general $\rho_f$, and thus the purity, $P$, will depend on 
whether or not the syndrome measurement detects an error or not. The purity 
of the final decoded state for both cases is shown in Fig.~\ref{QEC}. 

When $\alpha = \beta = 1/\sqrt{2}$ the QEC code perfectly corrects even complete 
dephasing on all three qubits depite the onset of tri-partite ESD. As $\alpha$ 
moves away from this point the purity decreases as a function of $p$, due to the 
inability of the QEC to correct the dephasing. Not surprisingly, some regions of low purity 
correspond to regions that exhibit ESD. However, even error regions in which the entanglement 
is detected by the entanglement witness can exhibit 
purity below $\approx .87$. This suggests that, for the particular QEC code studied 
here, any correlation between ESD and the success of the code 
is not fundamental. A comparison of the  onset of ESD due to the dephasing error with 
the purity is shown in Fig.~\ref{QEC}.

\section{Conclusions}
\label{Sconc}

In conclusion, I have studied the effect of dephasing on entanglement in three qubit 
systems. I have demonstrated ESD for both bi-partite and tri-partite 
entanglement using concurrence, the negativity, and the tri-partite negativity. These
were compared to the detection ability of tri-partite entanglement witnesses. 
In general we find that for mixed states the entanglement witnesses fail to 
detect the presence of entanglement well before the entanglement goes to zero. 
In addition, I have found a state in which there is no concurrence nor tri-partite 
negativity but entanglement as measured by the negativity still exists. 
Based on these exploration I considered whether ESD affects the workings of the 
three-qubit QEC phase flip code and concluded that there is no fundamental 
relationship between ESD and the failure of the code. 

As the number of system qubits grow studies of entanglement become more complex as the number of 
different types of entanglement continually increase. However,
a study of four qubit systems would allow for explorations of cluster state entanglement by 
using the proper entanglement witnesses \cite{TG}. This would be of central importance to issues 
of cluster state quantum computation \cite{RB}.

It is a pleasure to thank G. Gilbert and L. Viola for helpful feedback 
and acknowledge support from the MITRE Technology Program under MTP grant \#07MSR205.


\end{document}